\documentstyle[11pt,newpasp,twoside,epsf]{article}
\def\edcomment#1{\iffalse\marginpar{\raggedright\sl#1\/}\else\relax\fi}
\reversemarginpar

\begin{document}
\title{Massive star clusters in dwarf starburst galaxies}
 \author{G\"oran \"Ostlin}
\affil{Institut d'Astrophysique de Paris, 98bis Boulevard Arago, F-750~14 Paris,
France}
\affil{Present address: Stockholm Observatory, SE-133~36 Saltsj\"obaden, Sweden}

\begin{abstract}
I will discuss the presence of massive star clusters in starburst galaxies
with an emphasis on low mass galaxies outside the local group. I will show
that such galaxies, with respect to their mass and luminosity, may be very rich 
in young luminous clusters.

\end{abstract}


\section{Introduction}

During the last decade, the study of young massive stellar clusters, sometimes 
referred to as super star clusters (SSCs) has seen rapid progress. From a few
known examples of ``blue populous clusters'' in the Large Magellanic Cloud 
(LMC) and SSCs in the star forming dwarf galaxies NGC~1705  (Melnick, Moles, 
\& Terlevich 1985) and NGC~1569 (Arp \& Sandage 1985), the numbers have 
grown, largely thanks to HST. SSCs have now been found in a variety of
different environments, as have globular clusters (GCs) which are {\it old} 
massive clusters. 
In this paper I will review the massive cluster content of, 
 in particular, low mass starburst galaxies like blue compact galaxies (BCGs).
A review of massive clusters in starbursts  naturally becomes biased 
towards young SSC like objects. However,  such galaxies may
indeed contain also rich populations of older clusters.

The study of SSCs in starburst galaxies gained momentum with the advent of
the HST. Ultraviolet imaging with the aberrated HST/FOC (e.g. Meurer et al. 
1995, Conti and Vacca 1994) of starbursts revealed that a significant 
fraction of the star formation activity took place in SSCs. Meurer et al. (1995) 
studied nine galaxies, finding SSCs in most of them. Several other starbursts 
have been imaged with the aberrated HST/FOC (e.g. Conti et al., unpublished) and 
SSCs are frequently encountered.
Optical imaging with the aberrated HST has also  discovered  SSCs in many low mass 
starburst galaxies  (e.g. Hunter et al. 1994). The greater sensitivity of WFPC2 
as compared to FOC has multiplied the number of detected clusters in ESO~338-IG04 
(\"Ostlin, Bergvall, \& R\"onnback 1998) and He~2-10 (Johnson et al. 2000).

In giant galaxies there seems to be several ways to form SSCs/GCs, e.g. mergers 
(see Miller 2000), and bars (e.g. Kristen et al. 1997) and circum-nuclear rings 
in spiral galaxies (e.g. Barth et al. 1995). There are even indications
of SSC formation in the discs of normal spirals (Larsen \& Richtler 1999).
In low mass galaxies some of these mechanisms, e.g. formation of 
SSCs in bar and resonant induced density enhancements, are not available.
Mergers are certainly producing SSCs in some low mass galaxies (e.g. ESO~338-IG04, ESO~350-IG38, 
ESO~185-IG13), but there might be other mechanisms too. The origin of active star formation
in galaxies like NGC~1569 and NGC~1705 are not yet well understood. There are also
dwarf stabursts which do not contain luminous SSCs (e.g. IC10, see Grebel 2000).

Cluster destruction mechanisms (e.g. due to tidal shocks) are weaker in low mass 
galaxies giving SSCs  a greater chance of survival. In ESO~338-IG04  the 
dominant GC population is $\sim 3 $ Gyr old, and this population alone is enough to 
classify the galaxy as GC rich in terms of specific frequency. 
Low mass galaxies are in general metal-poor (typically [O/H]$\sim -1$, Kunth \& 
\"Ostlin 2000), which make them suitable
for comparison with high redshift conditions and early GC formation.
Dwarf galaxies have certainly been important ingredients in the hierarchial 
buildup of massive galaxies. Another virtue is that  
internal extinction in general is small, which for instance makes age dating more secure.

Thus, dwarf starbursts are good places to investigate the formation and evolution
of massive star clusters. Even if one cannot be sure whether a SSC will evolve into a bona 
fide GC or dissolve,  a proto-GC must look very much like a SSC (Kennicutt \& Chu 1988)
and the collective formation of a few SSCs will, by necessity, be associated with a starburst.
Thus populations of GCs and survived SSCs trace former starbursts, and  may  be used 
to study the evolution of galaxies.  
For example, if the excess of faint blue galaxies seen in deep optical counts is due to
starbursts originating in merging  galaxies at intermediate redshift, one 
would expect these to form  significant numbers of SSCs/GCs which should be visible as 
intermediate age GC populations in local 
galaxies.

\section{SSC richness in galaxies}

The richness of GCs in galaxies is often parametrized by the specific frequency, $S_N$,  
which is the number of GCs divided by the host galaxy luminosity (Harris \& van den Bergh 1981).
If counting only luminous clusters ($M_V \le -11$) and relating to $M_{V,host}$, 
the total absolute $V$ magnitude of the host galaxy, one may define a specific frequency 
of  luminous SSCs:  $S_{11} = N_{11} \times 10^{0.4(M_{V,host} + 15)}$, where $N_{11}$ is 
the number of clusters with $M_V \le -11$. 
$S_{11}$ will be independent of the assumed distance to a galaxy, and will be biased 
towards young massive clusters. The luminosity limit excludes contamination by supergiant 
stars and makes it possible to compare galaxies at different distances studied
at different depth.  
 
In Table 1, $S_{11}$ is given for a selection of galaxies of different types, where $N_{11}$ can
be obtained with reasonable accuracy (in nearby galaxies like M82 the large 
angular extent makes it difficult to estimate $N_{11}$). The table is by no 
means complete (see e.g. Miller 2000, for more references).  The $N_{11}$ 
values do not 
take internal extinction 
in the galaxies into account, but since the same is true for $M_{V,host}$, 
this should be a second order effect. If the luminous SSCs follow
a power law luminosity function, $\phi (L) \propto L^\alpha$, with $\alpha=-2$, $S_{11}$ will be independent of
the internal extinction as long as $A_V$  is equal for the SSCs and the integrated
galaxy light. This is not allways true since young SSCs may suffer from high
local extinction. If  $\alpha < -2$,
internal extinction will effectively lower the observed $S_{11}$ values.
All values in Table 1 assume $H_0 =75$ km/s/Mpc.
The $M_{V,host}$ values were assembled from different sources including
both accurate CCD photometry and  photographic magnitudes. Many apparent total
V-magnitudes have been taken, or estimated, from NED. Thus there might be an inhomogeneity on
the 0.5 magnitude level (corresponding to 50\% uncertainty on $S_{11}$), which 
one should be aware of when comparing galaxies. 

Values $S_{11} \le 0.5$ are found for giant mergers, while the luminous BCGs 
discussed below have $S_{11} \ge 0.6$. 
There are indications that these galaxies are actually dwarf mergers.
The nearby star forming galaxies NGC~1569, NGC~1705 and NGC~1140, have 
intermediate $S_{11}$ values. For LMC, no cluster is bright enough to qualify
without correcting for internal extinction. Most nearby dwarf irregular (dI) 
galaxies with modest star formation do not contain luminous SSCs (see  Grebel 2000). 
In the mergers, the number of luminous SSCs decrease 
with the estimated age of the merger remnant as was found by Schweizer et al. (1996).

\begin{table}[h]
\caption{Specific frequency of luminous SSCs for a selection of galaxies ($H_0 =75$km/s/Mpc).
See text for more details. 
 }
\begin{tabular}{llrrll}
\tableline

Galaxy		& Type	&	$M_{V,host}$ & 	$N_{11}$	& $S_{11}$ & Source for $N_{11}$	\\

\tableline
ESO338-IG04	&BCG	&	-19.3	&	53		&  1.0	& \"Ostlin et al. 1998	\\
Mrk~930		&BCG	&	-19.4	&	$\sim$110 	&  1.9	& This paper 		\\
ESO185-IG13	&BCG	&	-19.7	&	$\sim$105	&  1.4	& This paper		\\
ESO350-IG38	&BCG	&	-20.4	&	$\sim$130	&  0.9	& This paper		\\
He2-10		&BCG	&	-18.3	&	12		&  0.6	& Johnson et al. 2000	\\
NGC~1705	&Irr/BCG&	-16.2	&	1		&  0.3	& O'Connell et al. 1994	\\
NGC~1569	&Irr/BCG&	-17.8	&	4		&  0.3	& O'Connell et al. 1994 \\
NGC~1140	&Irr	&	-19.2	&	6		&  0.1  & Hunter et al. 1994	\\	
%NGC~253	&SAB	&	-19.5	&	4		&  0.06	& Watson et al. 1996 	\\
%LMC		&Irr	&	-18.3	&	1		&  0.05 & O'Connell et al. 1995 \\
%M~82		&	&	-19.4	&	0		&  0	& O'Connell et al. 1995 \\
NGC~1741	&Merger	&	-20.9	&	$\sim$50	&  0.2	& Johnson et al. 1999	\\
NGC~7252	&Merger	&	-22.1	&	$\sim$60	&  0.1 	& Miller et al. 1997 	\\
NGC~4038/4039	&Merger	&	-22.1	&	$\sim$150	&  0.2 	& Whitmore et al. 1999 	\\
NGC~3921	&Merger	&	-21.9	&	11		&  0.02	& Schweizer et al. 1996 \\
NGC~3256	&Merger	&	-21.9	&	$\sim$280	&  0.5 	& Zepf et al. 1999 	\\
NGC~1275	&Merger	&	-22.6	&	$\sim$75	&  0.07 & Carlson et al. 1998	\\
\tableline
\tableline

\end{tabular}
\end{table}

\section{A few case studies}

\subsection{ESO~338-IG04 (= Tol~1924-416)}

Ground based imaging had revealed an overdensity of faint blobs around this
well known blue compact galaxy, which was the motivation behind 
HST/WFPC2 followup observations (\"Ostlin et al. 1998).
These observations revealed  a starburst region composed of numerous blue
star clusters and a swarm of surrounding GCs. In all, the number of star
clusters (after correction for contamination by foreground stars, supergiant
stars and background galaxies) is above 100 (completeness limit $M_V \approx -9$ ~to~ $-10$). 
Spectral synthesis modelling 
of $U,B,V,I$ colors indicated the presence of distinct peaks in the cluster
formation history. In addition to the ongoing event there are old ($\sim 10$ Gyr) 
GCs, and in particular a very prominent population of intermediate age
(2 to 5 Gyr old) GCs. This intermediate age population contains the most 
massive cluster candidates and among them  an object (no. 34 in the outer 
sample of \"Ostlin et al. 1998) with an
estimated mass in excess of $10^7 M_{\sun}$ (for a  variety of different IMFs). 
ESO~338-IG04 is an intrinsically luminous ($M_V=-19.3$) metal poor ([O/H]$=-1$)
BCG. The dynamics and morphology suggest that the galaxy is the product of a
dwarf galaxy merger  (\"Ostlin et al. 1999, 2000). The current
specific frequency, $S_N = 2$,  is predicted to rise to $S_N \ge 10$
in one Gyr as the starburst ceases and fades.  The  most luminous SSCs 
were found already by Meurer et al. (1995).

\begin{figure}[h]

\plotone{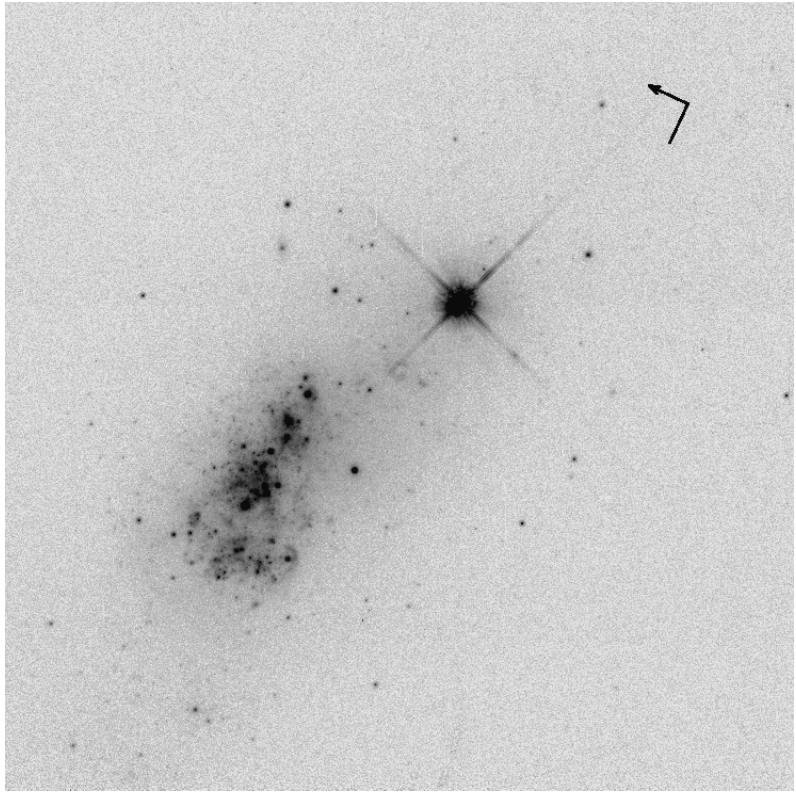}
%\plotfiddle{ostlin_fig1.eps}{12.5cm}{0}{75}{75}{-222}{-65}
\caption{\small ESO~338-IG04 imaged through filter F814W with 
the Planetary Camera of HST/WFPC2. The image size is $33\arcsec \times 
33\arcsec$. North is 
up-left (arrow head), east is down left. The length of the arrow is 1.8\arcsec,
corresponding to $\sim 330$ pc at a distance of 37 Mpc. 
 The bright source $\sim 5\arcsec$ west of the centre of the image is a foreground 
 star. }

\end{figure}

\subsection{He~2-10  (= ESO~495-G21)}

HST/FOC observations of He2-10 revealed ``knots'', some of which appeared 
to be resolved with diameters $< 10$ pc (Conti \& Vacca 1994). It was suggested
that these might be young GCs, but the use of a single UV passband  made it 
hard to say  more (e.g. addressing masses) than that the objects were young.
More recent WFPC2 observations (Johnson et al. 2000) confirm the FOC results, and 
multiply the detected number of clusters by reaching  fainter and redder objects. 
The faint, red objects may be intermediate age GCs or reddened SSCs. The galaxy
is one of the most nearby BCGs and a candidate dwarf merger.

\subsection{Markarian~996}

Markarian~996 is a blue compact dwarf ($M_V = -17.2$) with regular elliptical 
outer isophotes and intense star formation in a very compact central
H{\sc ii} region (Thuan, Izotov, \& Lipovetsky 1996). The central H{\sc ii} region
is bright enough to  saturate the WFPC2 images, but it could be a
luminous ($M_V \la -12$) young SSC. Intererestingly there are plenty of old GCs, 
asymetrically distributed around Mrk~996, with a luminosity function similar to 
that of Galactic GCs. Mrk~996 is, together with 
ESO~338-IG04, one of the rare known examples of a BCG posessing an old GC
population.  Mrk~996 has a GC specific frequency $S_N > 5$, similar to low-luminosity/dwarf
ellipticals (Miller et al. 1998).

\section{The Malkan et al. (1998) snapshot survey of AGN}

Malkan, Gorjian, \& Tam (1998) conducted an HST/WFPC2 snapshot survey of 
AGN, but included also a comparison sample consisting of 50 galaxies classified 
as having H{\sc ii}-activity. 
These ``H{\sc ii}'' galaxies can be divided in two broad classes: irregular galaxies
and spiral galaxies with nuclear star formation. 

Of galaxies having irregular or distorted morphology 3/4 appears to contain
at least a few SSCs, while only 1/3 of the spirals do. Galaxies classified as irregular
here are generally rather luminous and not typical dI galaxies like those in the 
local group. It is anyway obvious that the irregular galaxies, in view also 
of a lower average luminosity, appear to be  more efficient SSC formers than spirals.
Part of this effect could be explained by higher average extinction in spirals
than irregulars. The survey used only one pass band (F606W) and rather short 
exposures which were not cosmic ray split, making it of  limited use for quantitative 
studies. However, luminous SSCs can be easily identified and approximate photometry 
may be otained from  archive images.

The  survey includes ESO~185-IG13 and ESO~350-IG38 which
have been studied dynamically by \"Ostlin et al. (1999,2000). Both dynamics 
and morphology give strong support for a merger induced origin of the strong
starbursts seen in these galaxies, which  have similar properties to
ESO~338-IG04. These galaxies are also
among the most cluster rich in the whole survey.  Extracting photometry for the
SSCs in the HST archive images results in $S_{11}= 0.9$ and $S_{11}= 1.5$ for 
ESO~350-IG38 and ESO~185-IG13 respectively, see Table 1. ESO~350-IG38 contains
several SSCs with $M_V \sim -15$.

Another example, perhaps the best one in the whole survey, of a very SSC rich
starburst is Mrk~930, for which $S_{11} = 1.9$, see Table 1. Also Mrk~930 has 
a very irregular morphology. The Malkan et al.  (1998) survey also has a few objects in 
common with the study of Meurer et al. (1995).

\section{Conclusions and Perspectives}

Among dwarf and low mass galaxies we encounter both galaxies that appear to
be very efficient formers of massive clusters, and galaxies that appear totally
devoid of such objects. 

Some luminous BCGs have specific frequncies
of luminous ($M_V \le -11$) SSCs  $S_{11} \ge 1$, which is an order of magnitude
larger than most of the prototypical SSC factories: the ``Antennae'', NGC~7252 and NGC~1275.
Thus there is a tendency for  $S_{11}$  to increase  when going to low mass 
starburst  galaxies. A similar trend  has been found for the specific frequency of 
GCs among dwarf ellipticals (Miller et al. 1998). 
The purpose of comparing $S_{11}$ values was to show that low mass, metal-poor, 
starburst galaxies are excellent hunting grounds for luminous SSCs, and in addition 
problems with extinction are much smaller than in giant mergers. The BCGs with the highest
$S_{11}$ are believed to be the product of dwarf galaxy mergers (\"Ostlin et al. 1999, 2000). 
A possible explanation to the higher $S_{11}$ values is that the starburst timescales are
shorter in systems with lower mass, whereas in a giant merger one expects a more 
extended starburst. There are also BCGs and low mass starbursts which do not contain
luminous SSCs. Although the expected number of SSCs in galaxies of very low luminosity 
will always be small and subject to statistical fluctuations, there migh be a connection to
the triggering mechanism of the starbursts. Merging dwarfs might be more efficient SSC 
formers than non-merging ones.

A couple of BCGs (ESO~338-IG04 and Mrk~996)  in addition contains rich populations of older GCs. 
There is no a priori reason to believe that old GC systems are intrinsically rare among BCGs. 
Rather few BCGs have been studied at sufficient depth and spatial
resolution to unveil faint old GCs. The properties of relatively old GCs in BCGs
may provide important information about the nature of the host galaxy.

An unbiased survey of star forming dwarf galaxies with HST to characterize the
frequency of star clusters, their colors and host galaxy properties, would allow
to quantitatively study cluster formation in low mass galaxies.  The Malkan 
et al. (1998) survey do not fullfil these criteria but show that such a program
could be very rewarding.  A better understanding of the ultimate fate of SSCs is also
required. Dynamical mass
estimates are still rare and often result in masses of the right order of magnitude
but surprisingly low mass to light ratios (see  Smith \& Gallagher 2000). If this is 
due to flat or 
top heavy IMFs it would make it harder for young SSCs to survive and become GCs.

\begin{acknowledgements}
I would like to thank Kelsey Johnson and her collaborators for allowing me to 
discuss their work on He2-10 ahead of publication. I am also grateful to the 
organizers for arranging such a nice workshop. This work made frequent use of
the NASA/IPAC extragalactic database (NED).

\end{acknowledgements}

\section*{Discussion} 

%\vspace{0.1cm}

{\noindent \bf U. Fritze von Alvensleben:} I wonder if the star clusters you see in dwarf
galaxies could really be young globular clusters. Burst strengths and star formation efficiencies 
found for samples of blue compact dwarfs are $\la$ 1 \%, whereas GC formation modelling by Brown 
et al. (1995, ApJ 440, 666) requires star formation efficiencies of $\ge 10 $\%.  \\

{\noindent \bf G. \"Ostlin:} What matters is really the local star formation efficiency (SFE)
and not e.g. how large a fraction of the gas is converted into stars on a global
scale. Moreover, the SFE limit ($\ge 10 $\%) you refer to is the one required for a young GC 
to survive the Galactic tidal field, however tidal fields are  weaker in low mass galaxies. 
But of course, formation of numerous young 
globular clusters would require high 
global SF efficiencies and burst strengths. This is also the case for
the examples I am discussing, which all are galaxies with strong starbursts  (i.e. the gas depletion timescale, and/or the timescale to build up the observed stellar
mass is much shorter than a Hubble time). It is true that many galaxies termed 
``blue compact'' in reality have very modest star formation rates, but it is not in these
galaxies we predominantly find young GC candidates.  \\

{\noindent \bf J. Gallagher:} If most dwarfs make SSCs in reasonable numbers, why do we then see so 
few in WFPC2 images of ``normal'' dI galaxies? - Isn't this a problem? \\

{\noindent \bf G. \"Ostlin:} Normal dIs have mostly ``normal'' star formation rates 
with respect to the mass of the galaxy and are not experiencing starbursts 
(in terms of star formation timescales, see previous comment). Moreover, their luminosities are often 
so low that we would not expect many SSCs  if they had the same 
specific frequency of SSCs as e.g. ESO338-IG04 or giant mergers. Nevertheless,
it may be that formation of bound clusters require special conditions
(like high pressure or gas densities) which are not fulfilled in dwarfs
except in the case of external triggers, e.g. mergers and strong interactions.
The SSC rich dwarfs discussed above have perturbed morphology and signs of
mergers and/or interactions. \\

\end{document}